# Elastic Bound State in the Continuum with Perfect Mode Conversion

Liyun Cao[1,2], Yifan Zhu[2], Yanlong Xu[1], Shi-Wang Fan[2], Zhichun Yang[1,*], and Badreddine Assouar[2,*]

[1]*School of Aeronautics, Northwestern Polytechnical University, Xi'an, 710072, China*

[2]*Institut Jean Lamour, CNRS, Université de Lorraine, Nancy, 54506, France*

*Corresponding author e-mails: yangzc@nwpu.edu.cn & badreddine.assouar@univ-lorraine.fr

**Abstract**

The partial or complete confinement of waves in an open system is omnipresent in nature and in wave-based materials and technology. Here, we theoretically analyze and experimentally observe the formation of a trapped mode with perfect mode conversion (TMPC) between flexural waves and longitudinal waves, by achieving a quasi-bound state in the continuum (BIC) in an open elastic wave system. The latter allows a quasi-BIC in a semi-infinite background plate when Fano resonance hybridizes flexural and longitudinal waves and balances their radiative decay rates. We demonstrate that when the Fabry-Pérot resonance of the longitudinal wave is realized simultaneously, the TMPC formed by the elastic BIC approaches infinite quality factor. Furthermore, we show that quasi-BIC can be tuned continuously to BIC through the critical frequency of mode conversion, which offers the possibility of TMPC with an arbitrarily high quality factor. Our reported concept and physical mechanism open new routes to achieve perfect mode conversion with tunable high quality factor in elastic systems.

**Introduce**

Bound states in the continuum (BICs), defying the conventional bound states located outside the continuum, are non-decaying localized states embedded within the continuous spectrum of radiating waves (Hsu et al., 2016; Kodigala et al., 2017). The states can sometimes be regarded as embedded eigenvalues (Hsu et al., 2016) or embedded trapped modes (Hsu et al., 2013b) with infinite quality factors. The BIC concept was first proposed in quantum mechanics (von Neumann and Wigner, 1993) by mathematically constructing a 3D potential extending to infinity. Since this initial proposal, the wave phenomenon of BICs has been identified in different material and wave systems (Hsu et al., 2016), such as electromagnetic waves(Hsu et al., 2013b; Kodigala et al., 2017; Koshelev et al., 2019; Marinica et al., 2008; Minkov et al., 2018; Plotnik et al., 2011; Zhen et al., 2014), acoustic waves in the air (Cumpsty and Whitehead, 1971; Huang et al., 2020; Lyapina et al., 2018; Parker and Griffiths, 1968), water waves (Callan et al., 1991; Cobelli et al., 2011) and surface acoustic waves (Kawachi et al., 2001; Lim and Farnell, 1969; Trzupek and Zieliński, 2009) in semi-infinite media. Structures supporting BICs or quasi-BICs with high quality factor (high-Q) have been widely studied and opened the route to numerous applications in different fields, especially in optics and photonics, such as lasers (Hirose et al., 2014; Imada et al., 1999; Lin et al., 2020; Matsubara et al., 2008; Noda et al., 2001), sensors (Romano et al., 2019; Yanik et al., 2011; Zhen et al., 2013), and filters (Doskolovich et al., 2019; Foley et al., 2014; Ju et al., 2020). However, BICs in elastic waves systems are rarely recognized and concerned due to more sophisticated polarization states (Graff, 1975;

Rose, 1999) varying with the complex media structures.

On the other hand, perfect mode conversion (Giurgiutiu, 2007; Graff, 1975; Rose, 1999) between different polarization modes has grown into a burgeoning research area in the elastic wave system, due to its potential wide applications in nondestructive testing, medical ultrasonography, and earthquake resistance in civil engineering. For example, a mode-coupled layer with the balanced mode excitations and diagonal polarizations (Kweun et al., 2017) can achieve maximum mode conversion between longitudinal and shear modes, promoting the development of sensors in industrial and biomedical testing (Yang and Kim, 2018; Yang et al., 2019; Zheng et al., 2020). Pillared seismic metamaterials (Colombi et al., 2016a; Colombi et al., 2016b; Colquitt et al., 2017) can create effective bandgaps to achieve mode conversion from surface waves to bulk waves, promoting civil structures against seismic risk. Here, we propose a theoretical method to achieve perfect mode conversion between flexural waves (out-plane vibration) and longitudinal waves (in-plane vibration) (Graff, 1975; Rose, 1999) in two perpendicular polarization planes, which has never been reported. The new elastic-wave mode conversion mechanism enriches the form of elastic wave energy flow, which can perfectly convert an in-plane test signal or in-plane vibration energy into an out-plane one for the simplifications of nondestructive testing or energy harvesting, perfectly convert an out of plane vibration into in-plane one for reducing vibration.

In the present research, we present theoretical analysis and experimental evidence of achieving a trapped mode with perfect mode conversion (TMPC) in a quasi-BIC-supporting elastic wave system. In the proposed scheme, Fano resonance hybridizes flexural and longitudinal waves in two perpendicular polarization planes and converts one mode into another. When the radiative decay rate of the converted mode balances that of the directly reflected wave from the incident wave entering the system, the incident wave will be completely trapped by the hybrid Fano resonance, i.e., quasi-BIC. The resonant system only allows the converted mode to leak, leading to perfect mode conversion. Furthermore, we prove that TMPC can support infinite-Q BIC by simultaneously achieving the hybrid Fano and Fabry-Pérot resonances. We demonstrate that all quasi-BICs can be continuously tuned to BICs when the critical frequency of the mode conversion, depending on the incident angle, approaches the Fano resonance one.

**2. Theory and results**

To clearly show the unique TMPC in the quasi-BIC-supporting and BIC-supporting elastic wave systems, we investigate a simple structure with typical resonance characteristics. The structure is composed of a periodic waveguide resonator with the length $s$ on the edge of a background plate with infinite left boundary, as shown in Fig. 1(a). These waveguide resonators are connected to side-coupled pillared resonators with the height $h$. The thicknesses $d$ of the waveguide resonator and pillared resonator are in deep subwavelength scale relative to the wavelength in the whole considered frequency range. Only the fundamental modes of flexural waves and longitudinal waves are involved.

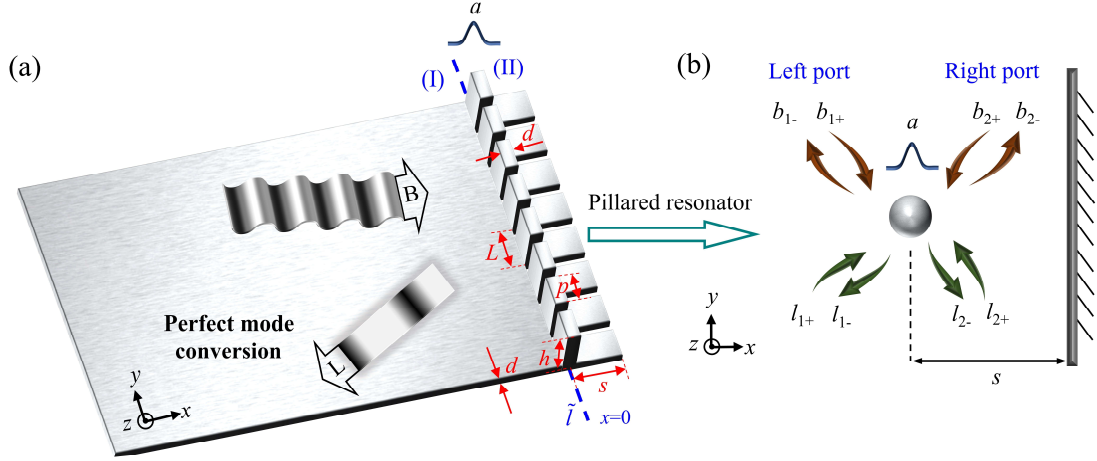

Fig. 1. (a) A Fano-resonance-supporting elastic wave model: periodic waveguide with the length $s$ connected to a side-coupled pillared resonator with the height $h$ on the edge of a semi-infinite background plate with the thickness $d$. The Fano resonance hybridizes flexural wave (bending wave) "B" and longitudinal wave "L" with polarizations in the $x$-$z$ plane and $x$-$y$ plane (perpendicular one), respectively. The resonance system perfectly converts one mode into another. (b) Schematic illustrations for the set-up of temporal coupled-mode theory in a subunit. The localized mode has amplitude $a$, which hybridizes the incoming flexural waves with amplitudes $b_{1+}$, $b_{2+}$, incoming longitudinal waves with amplitudes $l_{1+}$, $l_{2+}$, outgoing flexural waves with amplitudes $b_{1-}$, $b_{2-}$, and outgoing longitudinal waves with amplitudes $l_{1-}$, $l_{2-}$, respectively.

2.1. Analytical model of side-coupled pillared resonators with the Fano resonance and the hybridization coupling

As illustrated in the schematic of Fig. 1(b), the pillared resonator can hybridize the flexural and longitudinal waves in the background plate. First, without considering the right boundary, we get the scattering equation of the pillared resonator as

$$\mathbf{S}^- = \begin{bmatrix} t_b & r_b & t_{lb} & r_{lb} \\ r_b & t_b & r_{lb} & t_{lb} \\ t_{bl} & r_{bl} & t_l & r_l \\ r_{bl} & t_{bl} & r_l & t_l \end{bmatrix} \mathbf{S}^+ = \mathbf{M}_\mathrm{P} \cdot \mathbf{S}^+, \tag{1}$$

where $\mathbf{S}^- = [b_{2-}, b_{1-}, l_{2-}, l_{1-}]^T$ and $\mathbf{S}^+ = [b_{1+}, b_{2+}, l_{1+}, l_{2+}]^T$. $\mathbf{M}_\mathrm{P}$ is the scattering matrix. The symbols $t$ and $r$ denote the transmission and reflection coefficients, respectively. The subscripts $b$ and $l$ represent the flexural wave (or bending wave) and longitudinal wave. The subscripts $bl$ and $lb$ represent the conversion from flexural wave to longitudinal wave and the reverse process, respectively. Analytical expressions of all scattering coefficients in the scattering matrix $\mathbf{M}_\mathrm{P}$ are obtained by the transfer matrix method (see detailed derivation in Appendix A).

In the following, we study the case that the pillared resonator and the waveguide resonator have the same flexural rigidity, i.e., the same thickness, providing a strong hybridization coupling (Colquitt et al., 2017). The material parameters of the model are obtained based on the experimental measurement of 3D printing material PLA (see Appendix F). Figs. 2(a) and 2(b) display the dimensionless scattering energy curves for the incident flexural wave, based on the scattering coefficients in Eq. (1) (see details in Appendix A) and the full wavefield simulations, respectively. This consistency between Figs. 2(a) and 2(b) verifies the correctness of the analytical model. Another important verification is the power flow balance, i.e., the total energy of all scattering modes $e_t$ is unit (grey dot line). Note that the reflection energy curve $e_{r_b}$ has a clear Fano profile. The Fano profile is a typical feature of the Fano resonance (Fano, 1961; Rupin et al., 2014), which is the result of the interference between the flexural resonances of the pillared resonator and the incident flexural waves. The Fano resonance frequency $\omega_0$ [marked by yellow dashed lines in Fig. 2(a)], dominated by the flexural resonance, approximately satisfies the transcendental equation (Colquitt et al., 2017; Graff, 1975)

$$\cos\left(h\beta\sqrt{\omega_0}\right)\cdot\cosh\left(h\beta\sqrt{\omega_0}\right)=-1, \qquad (2)$$

where $\beta$ is the propagation constant, $\beta^4 = \rho d/D$, and $D$ is the bending rigidity.

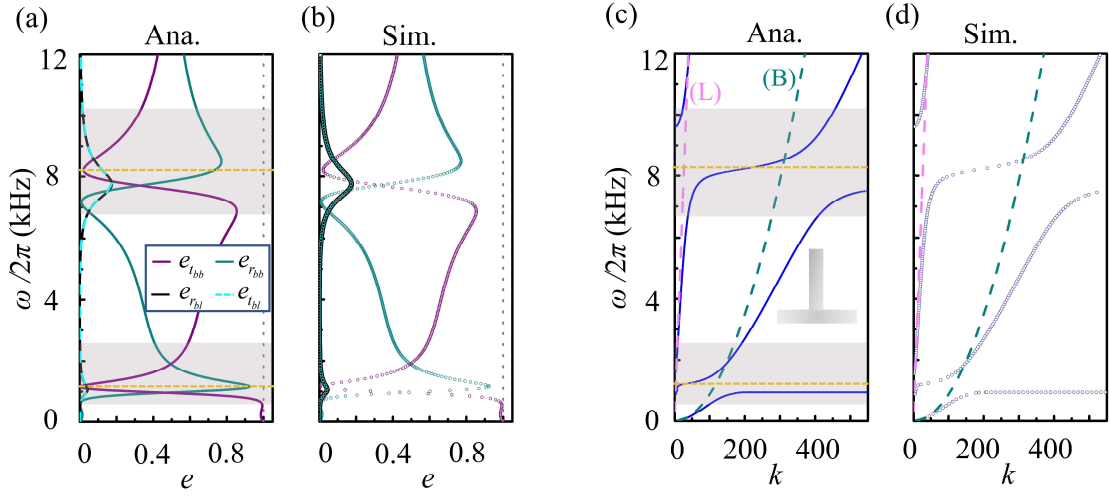

Fig. 2. (a) and (b) The dimensionless scattering energy curves for the incident flexural wave in the background plate without the boundary, based on the analytical and simulated methods, respectively. The grey dot line represents the total energy of all scattering modes. The yellow dashed lines indicate Fano resonance frequency. (c) and (d) Frequency–wavenumber representations correspond to the band structures for the pillared resonator, based on the analytical and simulated methods, respectively. The dispersion curves of the flexural wave (green dashed line) and longitudinal one (pink dashed line) in the background plate are superimposed.

Based on the analytical model, we analyze its band structure to intuitively show the hybridization coupling between flexural waves and longitudinal waves from the Fano resonance. Fig. 2(c) displays the band structure (blue lines), which are very consistent with the simulated one in Fig. 2(d). The dispersion curves of the flexural wave (green dashed line) and longitudinal wave (pink dashed line) in the background plate are superimposed. Around marked first and second Fano resonances [gray area in Fig. 2(b)], the band structures have typical three hybridized branches (blue lines): the right lower and left upper ones corresponding to the hybridized flexural and longitudinal modes, respectively; and the one in-between corresponding to the third mode of mixed symmetry (Rupin et al., 2014; Rupin and Roux, 2017). The three hybridized branches show strong mode coupling between flexural waves and longitudinal waves around Fano resonances.

2.2. The destructive interference relation based on the waveguide resonator

The introduction of the right waveguide resonator with the length $s$, as shown in Fig. 1, reshapes the interference of the incident flexural waves. The Fano resonance leaks the left scattering flexural waves into the background plate (channel one), and the right scattering flexural waves into the waveguide resonator. The right boundary of the waveguide resonator reflects all the right-going flexural waves, and part of that is transmitted through the pillared resonator into the background plate (channel two), interfering with channel one. Destructive interference can eliminate the radiation of flexural waves in these two channels to emerge a local state.

In the above intuitive understanding of the physical origin of the destructive interference, the coupling between the scattering modes of the resonance is an essential element. For theoretically explore the destructive interference in the elastic wave system, we apply temporal coupled-mode theory (Fan et al., 2003; Hsu et al., 2013a;

Hsu et al., 2013b) in optics, which provides a simple analytical description for resonant objects weakly coupled to incoming and outgoing propagation modes. As shown in Fig. 1(b), the amplitude $a$ of the localized mode evolves as $da/dt = (-i\omega_0 - \gamma_0)a$ in the absence of input powers. $\gamma_0$ is radiative decay rate of the localized mode. We can get the analytical expression of the destructive interference relation as [see details in Appendix]

$$e^{-i\phi} = t_b + r_b, \qquad (3)$$

where $\phi = 2k_b s - 3\pi/2$ is a phase shift of one round-trip of flexural waves in the waveguide, and $k_b$ is the wavenumber of the flexural wave. On the other hand, we can obtain the amplitude relation of the total scattering field $r_b + t_b$ as [see details in Appendix]

$$|r_b + t_b|^2 = 1. \qquad (4)$$

In such case, the magnitude condition of Eq. (3) is always satisfied and Eq. (3) can be reduced as its phase relation:

$$2\beta\sqrt{\omega}s - \frac{3\pi}{2} = \arg(r_b + t_b) + 2n\pi, \qquad (5)$$

where $n$ is a positive integer.

We should point out that hybridization coupling is an essential component of these unique local states from the destructive interference. The reason is that the wave fields in the system only have one mode without mode hybridization, then any state must be leaky. The radiative energy from mode conversion in our system is similar to dissipative loss in the optical resonance system (Hsu et al., 2016). But the "dissipative loss" in our study can be adjusted by the waveguide resonator, which will be discussed in detail later.

2.3. Analytical model of the whole resonator system

To evaluate all possible localized sates in the parameter space, we establish the analytical model of the whole resonator system. We can get the scattering equation of the whole resonator system as [see details in Appendix]

$$\begin{bmatrix} \left(R_b \quad R_b^* \quad R_{bl}\right)^{\mathrm{T}} \\ \mathbf{T}^{(\mathrm{II})} \end{bmatrix} = \mathbf{M}_1^{-1} \cdot \mathbf{G}_2 = \mathbf{M}_s, \tag{6}$$

where $\mathbf{M}_1 = \left(\mathbf{G}_1 \quad \mathbf{Q}_7\right)^{\mathrm{T}}$.

Based on Eq. (6), the dimensionless reflected energy coefficient $|R_b|^2$ of the vertically incident flexural wave, varying the waveguide resonator length $s/d$ and frequency $\omega$, is shown in Fig. 3(a). The scattering equation is the basis of the following theoretical study of elastic wave BIC, so it is necessary to verify its correctness by full-wavefield simulation. The simulated $|R_b|^2$ is shown in Fig. 3(b). Due to high computationally expensive of the simulation, we only used a large step for these variables of $s/d$ and $\omega$. However, we can still see that the overall trend of the results in Figs. 3(a) and 3(b) is the same. In addition, for a small range shown in the blue box in Fig. 3(b), these results are based on a fine variable grid and basically consistent with those in Fig. 3(a). For intuitively displaying the consistent, we chose the results at lines 1 and 2 in Fig. 3(b) to compare with that in Fig. 3(a), as shown in Fig. 3(c). According to Eqs. (6), (A18) and (A19), the total energy of the scattering modes $e_{\mathrm{T}}$ is calculated and shown in Fig. 3(d). The total energy $e_{\mathrm{T}}$ is unit, i.e., the power flow balance, which also verifies the correctness of the analytical model and indicates there is no other mode conversion, such as the SH wave, in our system.

2.4. Elastic BIC and elastic quasi-BIC physics

As shown in Fig. 3(a), the white dashed lines and green dotted lines, obtained by Eq. (2) and Eq.(5), correspond to the Fano resonance and destructive interference, respectively. Their intersections accurately predict the location of the vanishing linewidths, marked by the white circle. These vanishing linewidths are ideal local states with zero leakage in the continuous spectrum of radiating waves, i.e., the ideal BICs with infinite Q factor (Kodigala et al., 2017; Marinica et al., 2008).

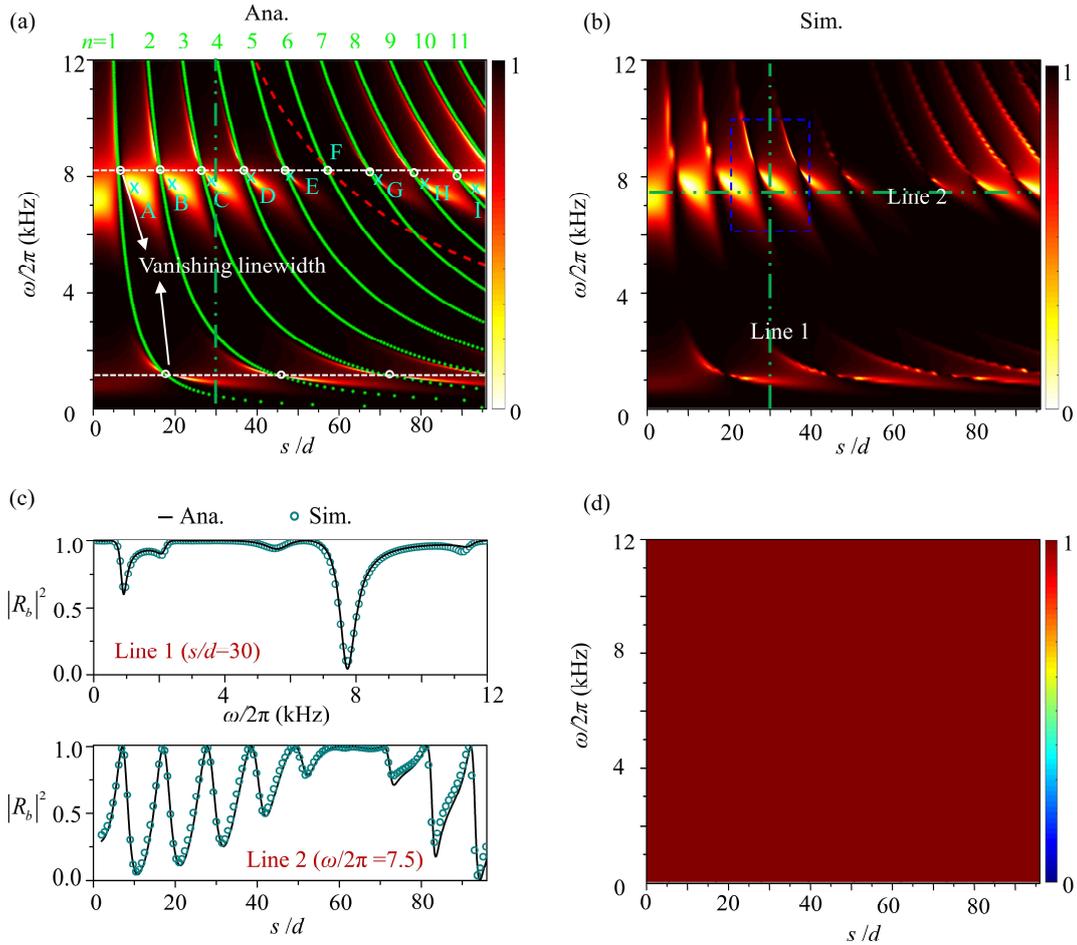

Fig. 3. (a) The analytical dimensionless reflected energy $|R_b|^2$ of the incident flexural wave in the whole resonant system, varying with frequency $\omega$ and the dimensionless waveguide length $s/d$. The

green dotted lines and white dashed lines correspond to the destructive interference and Fano resonance, respectively. White circles and green crosses indicate the vanishing linewidths and trapped modes, respectively. (b) The simulated dimensionless reflected energy $|R_b|^2$ by using a large step for these variables of $s/d$ and $\omega$. In the blue box, these results are based on a fine variable grid. (c) Comparison between the analytical solutions and the simulated ones at lines 1 and 2 in (b). (d) The total dimensionless energy of the scattering modes in the whole resonance system.

These ideal BICs at the vanishing linewidths are entirely isolated and have no access to the external radiation channel (Hsu et al., 2016; Huang et al., 2020). To get the TMPC, we need to tune the waveguide resonator length $s$ to weaken the destructive interference described by Eq. (5). This weakening makes the resonance system leaking small energy of the flexural wave (with radiative decay rate $\gamma_b$) radiating into the background plate. The leaked energy of the flexural wave will balance that of the longitudinal wave (with the radiative decay rate $\gamma_l$) from hybridization coupling of Fano resonance. This balance condition supporting trapped modes is $\gamma_b = \gamma_l$, equivalent to the so-called critical coupling (Cai et al., 2000), however here it is fulfilled by different mode conversion physics. The critical coupling can be demonstrated by the distribution of $\log_{10}|R_b|$ in the complex frequency plane, based on Eq. (6) (the detail see the Appendix B). In this way, the incident flexural waves can be entirely trapped without any backscattering ($|R_b|^2 = 0$), i.e., tuning into the trapped mode (quasi-BIC). In addition, the system only allows the converted longitudinal waves with a low radiative decay rate to leak, leading to perfect mode conversion. Here, we only analyze these TMPCs near the second Fano resonance, which are marked from point A to point I [see discrete green crosses in Fig. 3(a)]. Among these points, the position deviation between point A and the adjacent vanishing linewidth is the largest. The reason is that the radiative energy of longitudinal wave around point A is the largest (explained in the next paragraph), which needs more weakening of the destructive interference to leak more flexural wave energy to balance. The more leakage energy leads to a wider reflection band.

For these marked TMPCs, the total radiative decay rate of the system $\gamma = \gamma_b + \gamma_l = 2\gamma_l$ corresponds to the Q factor of the system $Q = \bar{\omega}_0/2\gamma = \bar{\omega}_0/4\gamma_l$, where $\bar{\omega}_0$ is the second Fano resonance frequency. Therefore, $\gamma_l$ decides the Q factor. When one round-trip of the converted longitudinal wave between the right waveguide edge and the left one (the interface $\tilde{l}$) satisfy Fabry-Pérot resonance condition, $\gamma_l$ approach zero due to the resonance, which induce the TMPC with infinite Q factor. This Fabry-Pérot resonance condition is obtained by enforcing field relations at the waveguide resonator, and accordingly,

$$2sk_l = m\pi, \tag{7}$$

where $m$ is a positive integer. According to Eq. (7), we can obtain Fabry-Pérot resonance frequency as $f_F = \sqrt{3}m/(2d\beta^2 s)$. We calculate the first resonance frequency ($m = 1$) varying with waveguide lengths $s$, as shown by the red dashed line in Fig. 3(a). The red line approximately crosses over the vanishing linewidth near point F, which indicates that near the ideal local state, $\gamma_l$ approaches zero due to the Fabry-Pérot resonance. Therefore, the leakage flexural wave energy from the trapped mode, balanced with the leakage longitudinal wave energy, also approaches zero. In this way, the position of TMPC almost overlaps with that of the ideal local state, approaching the BIC. At point F, the system with TMPC has a very narrow reflection band, which cannot be resolved in the scale of Fig. 3(a).

Based on $R_b$ in the analytical model, we obtain the Q factor and $\gamma$ of these TMPCs to quantitatively analyze the resonance characteristics (the detailed calculation can be found in Appendix C). It can be seen from Fig. 4(a) that the Q factor at point F exceeds 8000 (indeed approaches infinite-Q BIC). The closer to point F the other TMPC is, the greater its Q factor is, and the lower its $\gamma$ is. We chose two TMPCs with different radiative decay rate $\gamma$, corresponding to point A ($\gamma = 0.06\bar{\omega}_0$) and point E ($\gamma = 0.004\bar{\omega}_0$), to calculate their dimensionless x-direction energy distributions in the unit structure by full wavefield simulations, as shown in the insets of Fig. 4(a). It can be clearly seen from the insets that the flexural wave energy is trapped at the top of the pillared resonator. In addition, for point E with a lower $\gamma$, the system captures more flexural wave energy (the multiple of the dimensionless energy enhancement arrives up to 282). The reason is that the lower $\gamma$ is, the more trapped flexural wave energy in the resonance system with a higher Q is, which is similar to sound confinement from acoustic quasi-BIC (Huang et al., 2020). Note that by optimizing the parameters to satisfy Eqs. (2), (5), and (7) simultaneously, the Q of the system can be further increased. Therefore, our system can support the elastic BIC and have an infinite Q. According to Eqs. (6), (A18) and (A19), we also show the cures of energy conversion ratio for these TMPCs (from point A to point F) to verify their perfect mode conversion, as shown in Fig. 4(b). All peak values of the cures approach one. For point F, the linewidth of the curve at its half-maximum is less than 1 Hz, which is consistent with its high Q.

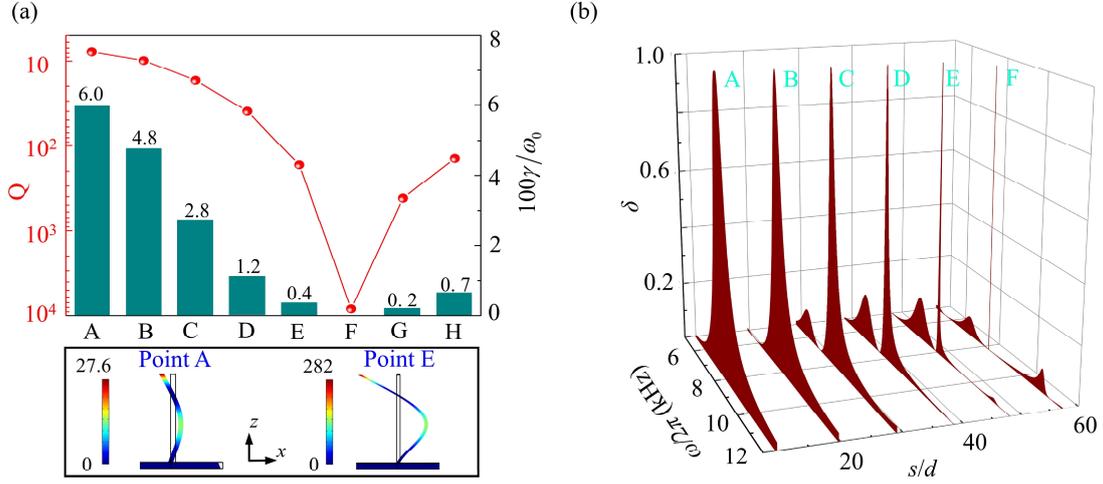

Fig. 4. (a) Bar charts represent total radiative decay rates $\gamma$ of these TMPCs from point A to point H in Fig. 3(a). Red balls correspond to Q of these TMPCs. (b) Dimensionless energy conversion spectra for these TMPCs.

2.5. Transforming elastic quasi-BIC to elastic BIC

As shown above, for vertically incident wave, adjusting the Fabry-Pérot resonance of the waveguide resonator by changing the waveguide length $s/d$, we can change $\gamma_l$ to obtain some discrete TMPCs with different Q, such as from point A to point I in Fig. 3(a). Further, for all these TMPC (quasi-BIC), we can continuously tune them into the infinite-Q BIC by making their radiative decay rate close to zero, depending on tuning the critical frequency of mode conversion.

The critical frequency is decided by that the radiation angle $\theta_{bl}^{r}$ of the converted longitudinal wave equal to 90º. The radiation angle can be obtained as $\theta_{bl}^{r} = \arcsin\left(k_b \sin\theta^i / k_l\right)$ based on Snell's law. From the above equation, we get the critical frequency as $f_c = 6\sin^2\theta^i / (\pi d^2 \beta^2)$, which is decided by the non-zero incident angle $\theta^i$. When the frequency is lower than $f_c$, the calculated $\theta_{bl}^{r}$ is imaginary, that is, the mode conversion disappears. For the TMPC marked by point B in Fig. 3(a), the dimensionless incident angle is changed as $\theta^i h/d = 77$, as an example. The interface impedance $Z_l = L/(p \cdot \cos\theta_{bl}^{r})$ with varying frequency is shown in Fig. 5(a). At the critical frequency of $f_c = 8384$ Hz marked by the light blue dashed lines, $Z_l$ is infinite, which will suppress longitudinal wave radiation. In addition, the corresponding $|R_b|^2$ for the obliquely incident flexural wave, calculated by Eq. (6), is shown in Fig. 5(b). The critical frequency closes to the Fano resonance frequency (marked by white dashed lines), the large $Z_l$ reduces $\gamma_l$ of the TMPC (marked by green crosses). The decreased $\gamma = 2\gamma_l$ leads to a narrower reflection band of the TMPC, compared with that of the TMPC marked by point B in Fig. 3(a).

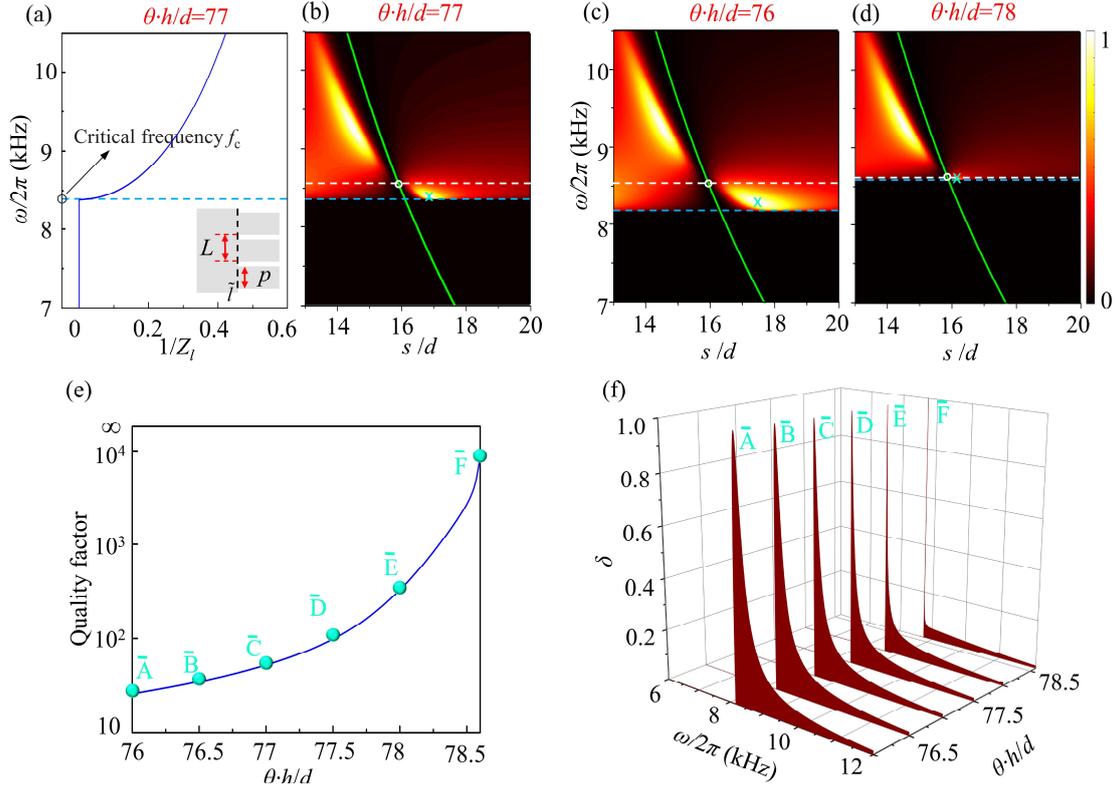

Fig. 5. (a) The interface impedance of the interface $\tilde{l}$ for the converted longitudinal wave with varying frequencies. (b)-(d) The reflected flexural energy coefficients of the whole system, in the frequency range from 7 kHz to 10.5 kHz and $s/d$ range from 13 to 20, when the incident angles are $\theta h/d = 77$, $\theta h/d = 76$, and $\theta h/d = 78$, respectively. The light blue dashed lines indicate the critical frequency of 8384 Hz, 8169 Hz, and 8603 Hz. (e) Q of the trapped mode can be continuously changed with the incident angle. Q is over 8700 at point $\bar{F}$. (f) The dimensionless energy conversion spectra of the structure at different incident angles. The linewidths of these curves at their half-maximum decrease from point $\bar{A}$ to point $\bar{F}$.

Figs. 5(c) and 5(d) show $|R_b|^2$ for two other incident angles. Comparing Figs. 5(b)-5(d), the closer to the Fano resonance frequency the critical frequency is, the narrower the reflection band of the system with TMPC is. The Fano resonance frequencies have a slight blueshift, because the vertical wave vector changes with different incident angles. In addition, the position of TMPC [marked by green crosses in Figs. 5(b)-5(d)] is closer to the local states (marked by white circles), due to the small radiative decay rate. In this way, we can continuously change the critical frequency position to change the Q of the TMPC, as shown in Fig. 5(e), where maximum Q is over 8700 at point $\bar{F}$. Fig. 5(f) shows peak values of dimensionless energy conversion spectra $\delta$ approach one for all TMPCs from point $\bar{A}$ to point $\bar{F}$. The spectra linewidths at their half-maximum gradually decrease with approaching point $\bar{F}$. Theoretically, when the critical frequency intersects the Fano resonance frequency, TMPC coincides with the zero-linewidth ideal local state and supports an infinite-Q BIC.

## 3. Experimental evidence

In the experiment, we need to make great efforts to excite the ideal plane wave through the optimized array composed of several piezoelectric patches. To simplify the excitation, we simplify the model in Fig. 1(a) by a strip-like model (beam), as shown in the inset of Fig.6(a). Then, we can use a single piezoelectric patch attached on the strip-like model to excite the ideal incident wave, which significantly increases the accuracy of the experiment. We have proved that for vertically incident waves, the results of the simplified model are consistent with those of the original model by theoretical and simulation methods (see details in Appendix D). To confirm our theory, based on the simplified models, we print specimen 1 and specimen 2 with different waveguide lengths to verify different-Q-factor TMPCs based on quasi-BIC. These TMPCs correspond to point A and point C in Fig. 3(a).

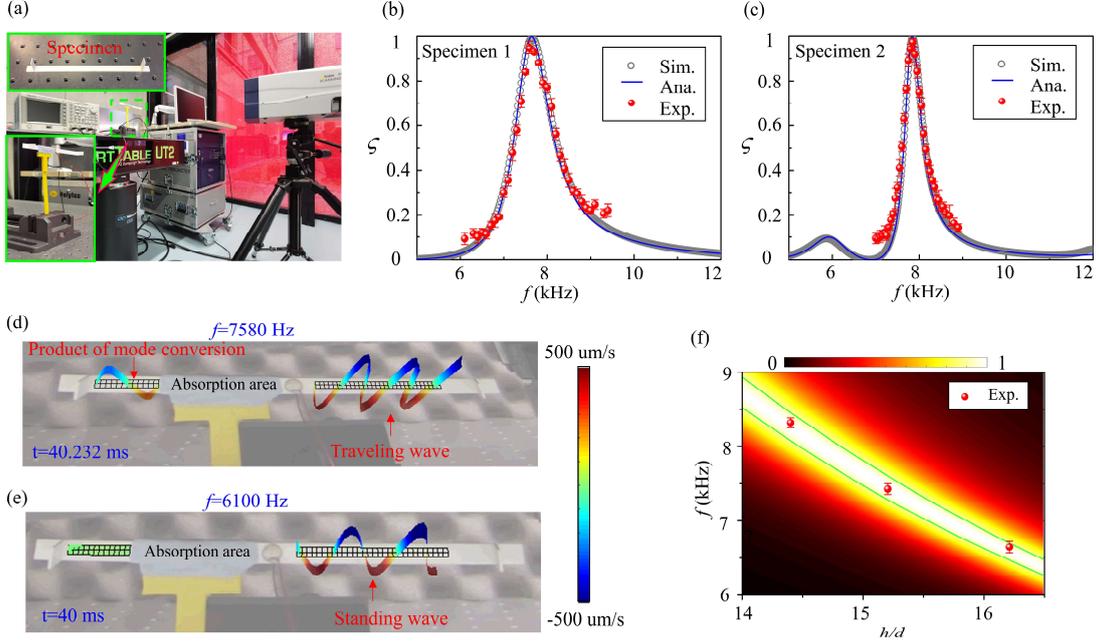

FIG. 6. (a) Experimental setup. The specimen and the clamped test specimen are illustrated at the top and bottom. (b) and (c) The capture coefficients based on the measured velocity fields of two simple specimens, respectively. Theoretical and numerical results are also displayed for comparative verification. (d) The snapshots of Supplementary Video 1 in the central frequency of 7580 Hz. (e) The snapshots of Supplementary Video 2 in the central frequency of 6100 Hz. (f) Theoretical capture coefficients, varying the pillared resonator height $h/d$ and frequency $f$. The waveguide resonator length is fixed. For three specimens with pillar heights $h/d$ of 14.2, 15, and 16, the locations of measured trapped modes are around 8516.7 Hz, 7629 Hz, and 6848 Hz, respectively.

We use a Polytec scanning vibrometer (PSV-500) to measure the out-of-plane velocity field in these specimens. Then we obtain the capture coefficient $\varsigma$ of flexural waves in these specimens. The detailed experimental measurements can be found in Appendix E. To ensure the reliability of the data, we tested each experimental data three times. The experiment results with error bars for specimen 1 and specimen 2 are shown by red balls in Figs. 6(b) and 6(c), respectively. It can be seen that the peak values at 7580 Hz and 7850 Hz are approximately one, that is, the flexural waves are almost completely captured. A series of simulations using COMSOL Multiphysics have also been performed. These capture coefficient curves from the experiments and simulations have excellent agreements with that from the theory [solid lines in Figs. 6(b) and 6(c)]. Therefore, we confirm the existence of the trapped modes with different Q factors. Furthermore, based on the experimental data, we calculate that the converted energy ratios of longitudinal waves are more than 0.93 and 0.9 for peak points in Figs. 6(b) and 6(c), respectively, which confirm the perfect mode conversion of the trapped modes, neglecting small structural losses. The detailed calculation process can be seen in Appendix G.

To intuitively show the TMPC, we obtain the dynamic full wavefield in specimen 1 at the central frequency of 7580 Hz by the ''time'' measurement mode of PSV 500. Fig. 6(d) shows a snapshot of Supplementary Video 1 at 40.232 ms. In the right test area, the incident wave is almost a traveling wave, whose intuitive evidence can be found in Supplementary Video 1. The existence of the traveling wave means flexural waves are completely captured by the right resonance system, that is, there is a very small reflection. The absorption area in the middle can completely absorb the flexural wave (relevant confirmation can be found in Appendix G), but we still clearly see the flexural wave field in the left test area. The abnormal phenomenon is caused by the mode conversion in the resonance system. For comparison, we test the dynamic full wavefield at 6100 Hz (away from the central frequency). Figs. 6(e) shows a snapshot of Supplementary Video 2. In the right test area, the incident wave is a standing wave. In the left test area, the amplitude of the flexural wave is close to zero. The intuitive evidence can be found in Supplementary Video 2. The wave phenomenon in Supplementary Video 2 means that flexural waves are not captured by the right resonance system, and mode conversion does not happen either.

We note a subtle but essential difference between the TMPC in our system, and the conventional trapped mode. The latter disappears when the parameters of the system slightly deviate from the designed structure, making it very difficult to observe experimentally (Hsu et al., 2013a; Hsu et al., 2016). In our theory, based on Eqs. (2) and (5), when the parameters of the system change, the vanishing linewidth jumps from one frequency to another. Therefore, the TMPC near the vanishing linewidth shift their positions and do not disappear. For example, theoretical capture coefficients, varying the pillared resonator height $h/d$ and frequency $f$, are shown in Fig. 6(f), when the waveguide length is fixed as 10 mm. The green lines are the contour lines of 0.9. High capture coefficients over a wide range indicate that the TMPCs are stable for various geometric parameters. Further, we have printed specimen 3 and specimen 4. Their pillar heights $h/d$ are 14.2 and 16. We can find the measured trapped modes ($\varsigma \approx 1$) for specimens 3 and 4 at around 8516.7 Hz and 6848 Hz, respectively. These measured points are added to Fig. 6(f). It can be seen that these points are inside the contour line of 0.9, which are consistent with the theory results. The consistency confirms that the TMPCs are robust to changes of system parameters. Similar robust of the trapped modes to system parameters can also be found in other BIC systems (Hsu et al., 2013a; Marinica et al., 2008).

## 4. Conclusions

We have theoretically predicted and experimentally demonstrated a TMPC in an elastic wave system. The system supports elastic BIC by achieving simultaneous hybrid Fano and Fabry-Pérot resonances. We prove that the TMPC supporting quasi-BIC can be tuned to the TMPC supporting infinite-Q BIC by bringing the critical frequency of mode conversion closer to the Fano resonance frequency. This work paves the way for the investigation of the intriguing physics related to the interaction between elastic wave energies with different polarized planes, based on elastic BICs. The elastic BIC can obtain an unprecedented high quality factor in the elastic system, which may facilitate various applications, e.g., high-sensitivity elastic wave sensor and high-resolution elastic wave filters.

**Acknowledgement**

The authors acknowledge Professor Yong Li for fruitful discussions. This project is supported by the Institute CARNOT (ICEEL), La Région Grand Est, the National Natural Science Foundation of China (Grant No.11972296), the 111 Project (No. BP0719007), and the Fundamental Research Funds for the Central Universities (No. 310201901A005).

# References

Bliokh, K.Y., Bliokh, Y.P., Freilikher, V., Savel'ev, S., Nori, F., 2008. Colloquium: Unusual resonators: Plasmonics, metamaterials, and random media. Rev Mod Phys 80, 1201-1213.

Cai, M., Painter, O., Vahala, K.J., 2000. Observation of Critical Coupling in a Fiber Taper to a Silica-Microsphere Whispering-Gallery Mode System. Phys Rev Lett 85, 74-77.

Callan, M., Linton, C.M., Evans, D.V., 1991. Trapped modes in two-dimensional waveguides. J Fluid Mech 229, 51-64.

Cao, L., Yang, Z., Xu, Y., Chen, Z., Zhu, Y., Fan, S.-W., Donda, K., Vincent, B., Assouar, B., 2021. Pillared elastic metasurface with constructive interference for flexural wave manipulation. Mech Syst Signal Pr 146, 107035.

Cao, L., Yang, Z., Xu, Y., Fan, S.-W., Zhu, Y., Chen, Z., Li, Y., Assouar, B., 2020. Flexural wave absorption by lossy gradient elastic metasurface. J Mech Phys Solids 143, 104052.

Cobelli, P.J., Pagneux, V., Maurel, A., Petitjeans, P., 2011. Experimental study on water-wave trapped modes. J Fluid Mech 666, 445-476.

Colombi, A., Colquitt, D., Roux, P., Guenneau, S., Craster, R.V., 2016a. A seismic metamaterial: The resonant metawedge. Sci Rep 6, 27717.

Colombi, A., Roux, P., Guenneau, S., Gueguen, P., Craster, R.V., 2016b. Forests as a natural seismic metamaterial: Rayleigh wave bandgaps induced by local resonances. Sci Rep 6, 19238.

Colquitt, D.J., Colombi, A., Craster, R.V., Roux, P., Guenneau, S.R.L., 2017. Seismic metasurfaces: Sub-wavelength resonators and Rayleigh wave interaction. J Mech Phys Solids 99, 379-393.

Cumpsty, N.A., Whitehead, D.S., 1971. The excitation of acoustic resonances by vortex shedding. J Sound Vib 18, 353-369.

Doskolovich, L.L., Bezus, E.A., Bykov, D.A., 2019. Integrated flat-top reflection filters operating near bound states in the continuum. Photonics Research 7, 1314.

Fan, S., Suh, W., Joannopoulos, J.D., 2003. Temporal coupled-mode theory for the Fano resonance in optical resonators. Journal of the Optical Society of America A 20, 569-572.

Fano, U., 1961. Effects of Configuration Interaction on Intensities and Phase Shifts. Phys Rev 124, 1866-1878.

Foley, J.M., Young, S.M., Phillips, J.D., 2014. Symmetry-protected mode coupling near normal incidence for narrow-band transmission filtering in a dielectric grating. Phys Rev B 89, 165111.

Giurgiutiu, V., 2007. Structural health monitoring with piezoelectric wafer active sensors.

Graff, K.F., 1975. Wave motion in elastic solids.

Hirose, K., Liang, Y., Kurosaka, Y., Watanabe, A., Sugiyama, T., Noda, S., 2014. Watt-class high-power, high-beam-quality photonic-crystal lasers. Nat Photonics 8, 406-411.

Hsu, C.W., Zhen, B., Chua, S.-L., Johnson, S.G., Joannopoulos, J.D., Soljačić, M., 2013a. Bloch surface eigenstates within the radiation continuum. Light: Science & Applications 2, e84-e84.

Hsu, C.W., Zhen, B., Lee, J., Chua, S.-L., Johnson, S.G., Joannopoulos, J.D., Soljačić, M., 2013b. Observation of trapped light within the radiation continuum. Nature 499, 188-191.

Hsu, C.W., Zhen, B., Stone, A.D., Joannopoulos, J.D., Soljacic, M., 2016. Bound states in the continuum. Nat. Rev. Mater. 1.

Huang, S., Liu, T., Zhou, Z., Wang, X., Zhu, J., Li, Y., 2020. Extreme Sound Confinement From Quasibound States in the Continuum. Phys Rev Appl 14.

Imada, M., Noda, S., Chutinan, A., Tokuda, T., Murata, M., Sasaki, G., 1999. Coherent two-dimensional lasing action in surface-emitting laser with triangular-lattice photonic crystal structure. Appl Phys Lett 75, 316-318.

Ju, C.-Y., Chou, M.-H., Chen, G.-Y., Chen, Y.-N., 2020. Optical quantum frequency filter based on generalized eigenstates. Opt Express 28, 17868-17880.

Kawachi, O., Mineyoshi, S., Endoh, G., Ueda, M., Ikata, O., Hashimoto, E., Yamaguchi, M., 2001. Optimal cut for leaky SAW on LiTaO/sub 3/ for high performance resonators and filters. IEEE Transactions on Ultrasonics, Ferroelectrics, and Frequency Control 48, 1442-1448.

Kodigala, A., Lepetit, T., Gu, Q., Bahari, B., Fainman, Y., Kanté, B., 2017. Lasing action from photonic bound states in continuum. Nature 541, 196-199.

Koshelev, K., Bogdanov, A., Kivshar, Y., 2019. Meta-optics and bound states in the continuum. Sci Bull 64, 836-842.

Krasnok, A., Baranov, D., Li, H., Miri, M.-A., Monticone, F., Alú, A., 2019. Anomalies in light scattering. Advances in Optics and Photonics 11.

Kweun, J.M., Lee, H.J., Oh, J.H., Seung, H.M., Kim, Y.Y., 2017. Transmodal Fabry-Perot Resonance: Theory and Realization with Elastic Metamaterials. Phys Rev Lett 118.

Lim, T.C., Farnell, G.W., 1969. Character of Pseudo Surface Waves on Anisotropic Crystals. The Journal of the Acoustical Society of America 45, 845-851.

Lin, F., Ullah, S., Yang, Q., 2020. Ultrafast vortex microlasers based on bounded states in the continuum. Sci Bull 65, 1519-1520.

Lyapina, A.A., Pilipchuk, A.S., Sadreev, A.F., 2018. Trapped modes in a non-axisymmetric cylindrical waveguide. J Sound Vib 421, 48-60.

Marinica, D.C., Borisov, A.G., Shabanov, S.V., 2008. Bound States in the continuum in photonics. Phys Rev Lett 100, 183902.

Matsubara, H., Yoshimoto, S., Saito, H., Jianglin, Y., Tanaka, Y., Noda, S., 2008. GaN Photonic-Crystal Surface-Emitting Laser at Blue-Violet Wavelengths. Science 319, 445-447.

Minkov, M., Williamson, I.A.D., Xiao, M., Fan, S., 2018. Zero-Index Bound States in the Continuum. Phys Rev Lett 121, 263901.

Noda, S., Yokoyama, M., Imada, M., Chutinan, A., Mochizuki, M., 2001. Polarization Mode Control of Two-Dimensional Photonic Crystal Laser by Unit Cell Structure Design. Science 293, 1123-1125.

Parker, R., Griffiths, W.M., 1968. Low frequency resonance effects in wake shedding from parallel plates. J Sound Vib 7, 371-379.

Plotnik, Y., Peleg, O., Dreisow, F., Heinrich, M., Nolte, S., Szameit, A., Segev, M., 2011. Experimental Observation of Optical Bound States in the Continuum. Phys Rev Lett 107, 183901.

Romano, S., Zito, G., Lara Yépez, S.N., Cabrini, S., Penzo, E., Coppola, G., Rendina, I., Mocellaark, V., 2019. Tuning the exponential sensitivity of a bound-state-in-continuum optical sensor. Opt Express 27, 18776-18786.

Rose, J.L., 1999. Ultrasonic waves in solid media. Cambridge University Press 10.

Rupin, M., Lemoult, F., Lerosey, G., Roux, P., 2014. Experimental demonstration of ordered and disordered multiresonant metamaterials for lamb waves. Phys Rev Lett 112.

Rupin, M., Roux, P., 2017. A multi-wave elastic metamaterial based on degenerate local resonances. J Acoust Soc Am 142, EL75.

Trzupek, D., Zieliński, P., 2009. Isolated True Surface Wave in a Radiative Band on a Surface of a Stressed Auxetic. Phys Rev Lett 103, 075504.

von Neumann, J., Wigner, E.P., 1993. Über merkwürdige diskrete Eigenwerte, in: Wightman, A.S. (Ed.), The Collected Works of Eugene Paul Wigner: Part A: The Scientific Papers. Springer Berlin Heidelberg, Berlin, Heidelberg, pp. 291-293.

Yang, X., Kim, Y.Y., 2018. Asymptotic theory of bimodal quarter-wave impedance matching for full mode-converting transmission. Phys Rev B 98, 144110.

Yang, X., Kweun, M., Kim, Y.Y., 2019. Monolayer metamaterial for full mode-converting transmission of elastic waves. Appl Phys Lett 115, 071901.

Yanik, A.A., Cetin, A.E., Huang, M., Artar, A., Mousavi, S.H., Khanikaev, A., Connor, J.H., Shvets, G., Altug, H., 2011. Seeing protein monolayers with naked eye through plasmonic Fano resonances. Proceedings of the National Academy of Sciences 108, 11784-11789.

Zhen, B., Chua, S.-L., Lee, J., Rodriguez, A.W., Liang, X., Johnson, S.G., Joannopoulos, J.D., Soljačić, M., Shapira, O., 2013. Enabling enhanced emission and low-threshold lasing of organic molecules using special Fano resonances of macroscopic photonic crystals. Proceedings of the National Academy of Sciences 110, 13711-13716.

Zhen, B., Hsu, C.W., Lu, L., Stone, A.D., Soljačić, M., 2014. Topological Nature of Optical Bound States in the Continuum. Phys Rev Lett 113, 257401.

Zheng, M., Park, C.I., Liu, X., Zhu, R., Hu, G., Kim, Y.Y., 2020. Non-resonant metasurface for broadband elastic wave mode splitting. Appl Phys Lett 116, 171903.